# Vortex motion rectification in Josephson junction arrays with a ratchet potential

D. E. Shalóm* and H. Pastoriza†

*Centro Atómico Bariloche, Comisión Nacional de Energía Atómica,*
*Av. Bustillo 9500, R84002AGP S. C. de Bariloche, Argentina*

(Dated: November 19, 2004)

By means of electrical transport measurements we have studied the rectified motion of vortices in ratchet potentials engineered on over-damped Josephson junction arrays. The rectified voltage as a function of the vortex density shows a maximum efficiency close a matching condition to the period of the ratchet potential indicating a collective vortex motion. Vortex current reversals where detected varying the driving force and vortex density revealing the influence of vortex-vortex interaction in the ratchet effect.

The rectification of motion by asymmetric periodic potentials, so called *ratchet effect*, has been addressed extensively in recent years [1–3]. The suggestion that directed motion in biological systems is driven by this effect have largely triggered the research in this subject [4]. In this letter we address this subject by studying superconducting vortices which has been proven to be a paradigm for testing a number of statistical phenomena due to the ability of controlling density and interactions. The ratchet effect on superconducting vortices has been studied for numerous vortex pinning potentials [5–8] and suggested as a method to reduce vortex density or even generate lensing and guidance of vortices [9, 10]. In Josephson junction arrays (JJA) and SQUIDs, asymmetric potentials for vortices and fluxons were also proposed and studied [11–13]. In this letter we present measurements on JJAs where asymmetric periodic pinning potential were created. We were able to detect rectification in the vortex motion and study the ratchet effect. By changing the periodicity of the potential we were able to study collective effects in this phenomena.

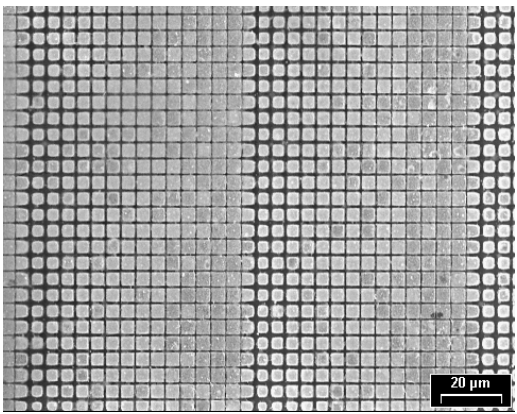

FIG. 1: Scanning electron microphotograph of a region of the sample V15. Light grey regions are Pb superconducting islands, and dark grey regions are Cu underneath. Lattice parameter is $5\mu$m, and junction gap varies from 0.2 to $1\mu$m.

Our JJAs were fabricated by e-beam lithography and $Ar^+$ ion milling starting from a 2000 Å/2000 Å Lead-Copper bilayer. The gap between Pb islands was modulated with a sawtooth function across the arrays from 0.2 to $1\mu$m, while keeping $5\mu$m as the center to center distance between islands. In all cases sample size were $100 \times 100$ islands. Different samples with periods of the ratchet potential $P = 7$, 10 and 15 array cells were built maintaining the overall width of the array constant (hereafter called V7, V10 and V15 respectively). In Fig. 1 we show a scanning microscope picture of a representative region of one of the samples.

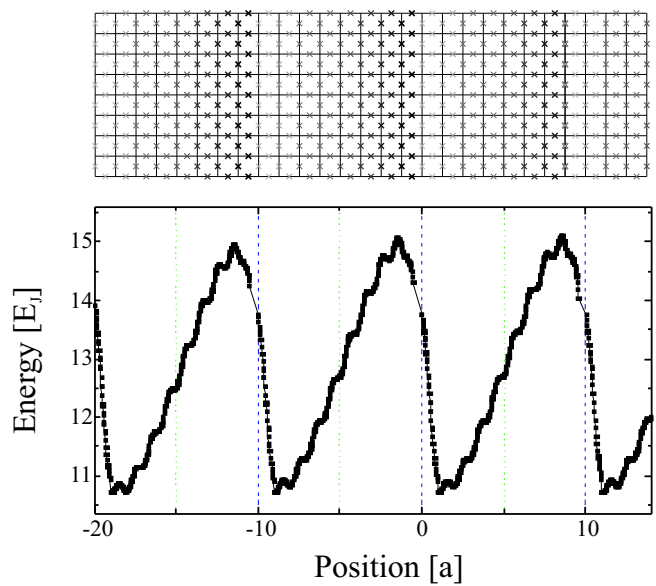

FIG. 2: Upper part: Scheme of the ratchet JJA with modulated coupling energy $E_J$. The crosses represent the Josephson junctions, with its width proportional to the magnitude of $E_J$. Lower part: Potential energy of a vortex in a ratchet JJA as a function of position, obtained by numerical simulations. Array size $N = 60$; $P = 10$; coupling energy $E_J$ varies linearly from 0.5 to 1.4.

Due to the discreteness of the JJA the energy associated with a single vortex is position dependent [14]. This feature is responsible for the existence of a finite critical force for the vortex motion (made evident through a critical current) and is responsible for the guiding of vortices [15]. To obtain an insight on the energy landscape for a vortex in these ratchet JJAs we performed a series of numerical simulations. We modeled a square array of superconducting islands connected by ideal Josephson junctions of coupling energy $E_J$. No resistive or capacitive terms were considered. Every junction energy can be summed to give the Hamiltonian

$$H = \sum_{<i,j>} E_J^{i,j}[1 - \cos(\varphi_i - \varphi_j - A_{ij})],$$

where $\varphi_i$ stands for the phase of the superconducting order parameter in the island $i$, the sum is taken over nearest-neighbor pairs, $A_{ij} = \frac{2\pi}{\Phi_0}\int_j^i \vec{A}d\vec{l}$, $\vec{A}$ is the vector potential, and $E_J^{i,j}$ embodies the modulation of the coupling energy to build the ratchet potential. This Hamiltonian assumes an infinite penetration depth of the magnetic field, $\lambda_\perp = \infty$ (no self fields are taken into account). Within this model the arctan aproximation has been used to construct the phase configurations of a vortex in a homogeneous JJA, and from there, the energy landscape is calculated [16]. It is difficult to generalize this atan procedure to an inhomogeneous JJA, where coupling energies vary from one junction to the other. For our ratchet arrays we used a variation of a numerical relaxation technique [16, 17] to calculate the potential energy of the vortex. First we introduce a vortex by using the arctan approximation as an initial condition. Then, we let the phases $\varphi_i$ to relax to the equilibrium configuration following a Metropolis algorithm. As a result we obtain the phase configuration and energy of a state corresponding to a vortex located on the bottom of a potential well [17]. To obtain the energy of states other than these local minima, we introduced a variation to the previous procedure [18], in which we fix and control the phase of two islands, adjacent to the vortex center, while allowing all other phases in the system to relax. In this way, we are able to pull and push the vortex uphill, as we are forcing the center of rotation of the vortex currents to be in a defined location, allowing us to calculate the potential energy of a vortex located in any arbitrary position. Using this method we have numerically calculated the single vortex potential for a linear sawtooth coupling energy dependence across the array, for zero magnetic field. The result of this calculation shows clearly the almost ideal ratchet landscape, as seen in lower part of Fig. 2, supporting the idea of using gap-modulated JJAs to create a ratchet potential.

The temperature dependence of the coupling energy for a superconductor-normal-superconductor (S/N/S) Josephson junction is well described by the de Gennes expression:

$$E_J(T) = \frac{\hbar}{2e}I_0(1 - T/T_0)^2 \exp(-d/\xi_N(T))$$

where $I_0$ represents the zero temperature critical current, $T_0$ the critical temperature of the superconducting electrodes, $d$ the distance between superconductors, and $\xi_N(T) = \xi_0/\sqrt{T}$ the coherence length in the normal metal. In our sample design, where the parametrized magnitude is $d$, the landscape of the vortex potential is temperature dependent. However, due to the dependence of the coupling energy with the gap between islands, the asymmetry of the *ratchet*-like potential is preserved for all non-zero temperatures.

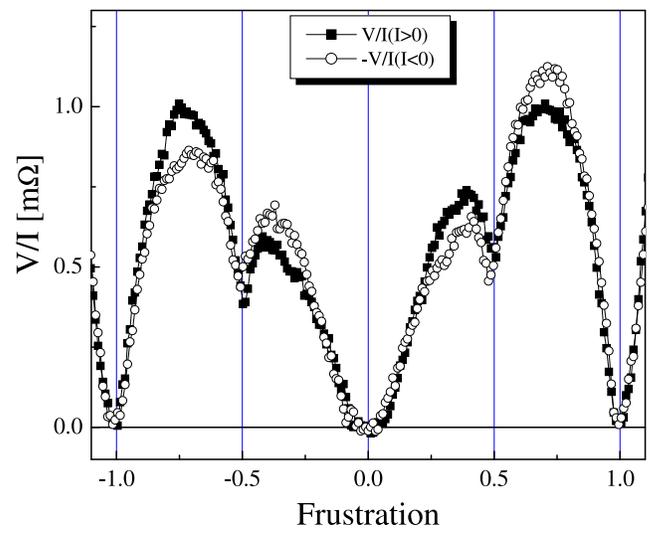

FIG. 3: V/I as a function of the frustration of the sample V10 for a temperature of $T = 3.2$ K and for positive and negative dc current $I = 0.4$ mA. The difference in the resistance between both current directions is evident, indicating the rectification of the vortex motion in the array.

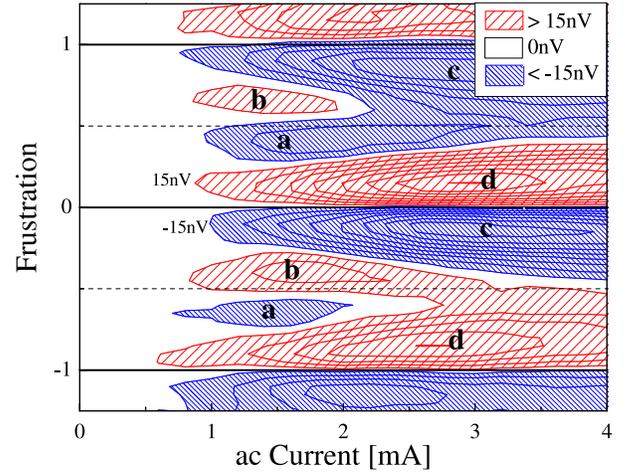

FIG. 4: Rectified voltage as a function of frustration and ac current for sample V7 at $T - 3.8$ K and $\nu = 1$ kHz. Regions dashed diagonally upper-left to lower-right (blue, dense) indicate negative voltage values. Regions dashed diagonally lower-left to upper-right (red, sparse) indicate positive voltage values. Contour lines are drawn every by 30 nV. Regions **c, d** correspond to matching states of densities $N \pm 1/P$ (see text).

In order to test the rectifying efficiency of the designed samples, we have performed I-V measurements with the force exerted by the current either forward or backward to the sawtooth potential. In Fig. 3 we show V/I data of the sample V10 as a function of the magnetic flux in the sample, for both current directions, for a temperature of 3.2 K and an applied current of 0.4 mA. The magnetic flux is expressed as frustration $f$, defined as the applied magnetic flux in units of flux quan-tum $\Phi_0$ per plaquette ($5 \times 5\,\mu m^2$). This plot reveals a number of interesting features. It is clear the periodicity of the data with integer numbers of frustration, and the minimum at $f = 1/2$, associated to symmetries of the Hamiltonian that describes a JJA [19]. This periodicity is convoluted by the single junction effect due to the design of our arrays where the area of a single junction is about 5-20% of the array cell area. The difference in the resistance between both current directions is evident from the plot, indicating the rectification of the vortex motion in the array. A distinguished result of this measurements is the change of sign in the difference between both resistances each half period. This is a direct consequence of the periodicity of the Hamiltonian with the magnetic field. In terms of vortex excitations it can be explained by considering the *sign* of the driven vortices: from $f = 0$ to $f = 1/2$ the vortex ensemble consists of a discrete number of *positive* vortices which will be dragged by the asymmetric potential toward the forward direction. On the other hand from $f = 1/2$ to $f = 1$ the much more dense vortex ensemble can be regarded as *negative* vortices in a fully filled static background of positive vortices. Both types of vortices have the same pinning landscape as it is mainly dominated by the geometry of the array. Therefore either positive and negative vortices are dragged toward the same direction by the ratchet potential. However the induced voltage from vortex motion depends on the vortex *sign*. This means that if a positive vortex moving to the *right* induces a positive voltage, a negative vortex moving to the *right* will generate a negative voltage. The mirrored effect happens for negative frustrations. From $f = 0$ to $f = -1/2$ the net motion is from an ensemble of negative vortices, and from $f = -1/2$ to $f = -1$ corresponds to positive vortices moving in a negative-vortex background. Although this picture for the vortex structures has been presented in previous works [20] to our knowledge this is the first experimental evidence for the *sign* of the mobile vortices.

We have also driven the system as a rocking ratchet[3], by applying an alternating current of varying amplitude and frequency while detecting the rectified dc induced voltage as a function of frustration or temperature. This is similar to make the subtraction of both data sets of Fig. 3 but with an overall increase in speed and the signal-to-noise ratio of the acquisition. In Fig. 4 we plot as two dimensional graph the magnitude of the rectified voltage as a function of the frustration and applied ac current for the sample V7 at a temperature of 3.8 K. A frequency of $\nu = 1\,kHz$ was used for the measurements presented here. Other frequencies were also explored, and the same results were obtained within the region of frequencies from 10 Hz to 100 kHz. Similar plots were obtained for samples V10, and V15 and other temperatures, always presenting equivalent features to those described hereafter. At low currents, $I < 1\,mA$ for data shown in Fig. 4, no voltage is detected because the ac current is lower than the critical current, and no vortex movement is generated. This fact was verified measuring V-I characteristics under the same conditions and observing that the onset of both voltages (rectified from the ac experiment and from the VI measurement) coincides.

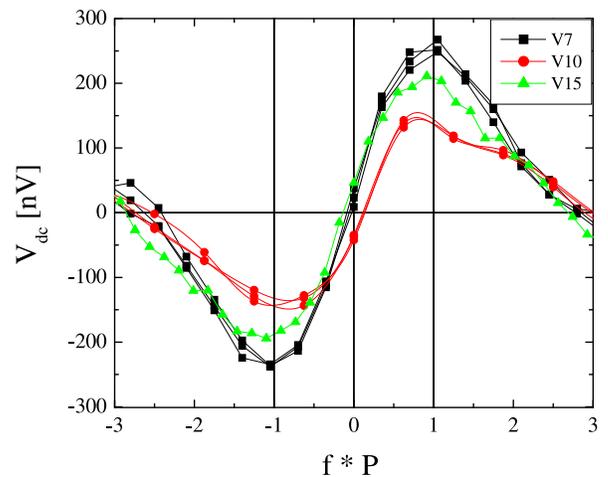

FIG. 5: Peak structure in the rectified voltage as a function of the frustration normalized by the inverse of the ratchet potential period $P$, for the three samples V7, V10 and V15. V7: $T = 3.8\,K$, $I = 3\,mA$. V10: $T = 2.6\,K$, $I = 2.1\,mA$. V15: $T = 2.8\,K$, $I = 5\,mA$.

At higher currents a periodic in field complex response is observed with voltage reversals as varying magnetic field (vortex density) and ac current (driving force).

At low vortex densities a pronounced structure is observed, for frustration $f \simeq \pm 1/7$ labeled **c** and **d** in Fig. 4. It is also seen at $f \simeq (n \pm 1/7)$, with $n = \pm 1$. Similar feature is found for the V10 and V15 samples with position of the maximum located in $f = 1/P$. This is clearly shown in Fig. 5 where the rectified voltaje is plotted as a function of frustration times $P$ for the three samples investigated. The applied ac current selected for each sample was chosen to cross the absolute maximum at the given temperature. For example, the sample V7 and $T = 3.8\,K$ we present data for $I_{ac} = 3\,mA$, as could be estimated from Fig. 4. The overlap of the curves is remarkable indicating that there is a optimal rectification of the vortex motion when there is a matching condition between the moving vortex structure and the periodic ratchet potential.

Based on the matching condition for a peak at $f_{max} = 1/P$ for a sample of period $P$ of the ratchet potential, we can especulate that the moving structure would be similar to a row of adjacent vortices located at the bottom of every tooth of the sawtooth potential. It is difficult to predict which is the vortex configuration as a function of frustration for such ratchet arrays. Therefore it is not trivial to describe which is the matched state responsible of the observed rectification and dynamical numerical simulations would be required to elucidate this issue.

A more complex interpretation is required to analyse the sign reversal observed at lower currents in the regions indicated as **a** and **b** in Fig. 4. The fact that this feature is observed for high densities can be taken as an indication that vortex-vortex interactions overcome the geometrical potential landscape and effectively change the shape and asymmetry of the ratchet potential. A change in the structure of the vortex system occurs on dynamical phases[21], changing the vortex-vortex interactions. As a result a sign reversal of the rectified motion can arise at high vortex velocities. More experiments complemented with dynamical simulations are required to give a more detailed explanation of these features.

In conclusion we have presented a design of Josephson junction arrays that generates a sawtooth-like potential for vortices and antivortices. Transport measurements indicate a rectified motion for these excitations. These experiments also clearly identify for the first time the *sign* of the transported vortices which changes as a function of frustration. Measurements performed in samples with different periodicities of the sawtooth potential show that there is a maximal rectification of the vortex motion for a matching condition of the vortex density to the ratchet periodicity. Our results indicate that collective effects are relevant for understanding the motion of particles in ratchet potentials. It was suggested [9] that the ratchet effect can be used to reduce the vortex density in superconductors. Our results widen this suggestion showing the possibility of the utilization of potentials with different periodicities for density separation.

We thank S. Reparaz for ideas and assistance in sample fabrication in the genesis of this work. We acknowledge useful discussions with D. Domínguez and F. de la Cruz.